# Capturing Near Earth Objects[*]

Hexi Baoyin, Yang Chen, Junfeng Li

Department of Aerospace Engineering, Tsinghua University 100084 Beijing, China

**Abstract:** Recently, Near Earth Objects (NEOs) have been attracting great attention, and thousands of NEOs have been found to date. This paper examines the NEOs' orbital dynamics using the framework of an accurate solar system model and a Sun-Earth-NEO three-body system when the NEOs are close to Earth to search for NEOs with low-energy orbits. It is possible for such an NEO to be temporarily captured by Earth; its orbit would thereby be changed and it would become an Earth-orbiting object after a small increase in its velocity. From the point of view of the Sun-Earth-NEO restricted three-body system, it is possible for an NEO whose Jacobian constant is slightly lower than $C_1$ and higher than $C_3$ to be temporarily captured by Earth. When such an NEO approaches Earth, it is possible to change its orbit energy to close up the zero velocity surface of the three-body system at point $L_1$ and make the NEO become a small satellite of the Earth. Some such NEOs were found; the best example only required a 410*m/s* increase in velocity.

## 1. Introduction

Some Jovian comets, such as Oterma, are sometimes temporarily captured by Jupiter, making the transition from heliocentric orbits outside the orbit of Jupiter to heliocentric orbits inside the orbit of Jupiter. During this transition, Jupiter frequently captures the comet temporarily for one to several orbits (Koon et al. 2000). This is because the Jacobian constant of the comet is slightly lower than $C_1$ (Jacobian constant at Sun-Jupiter Lagrange point $L_1$) and higher than $C_3$ (Jacobian constant at Sun-Jupiter

[*] This research is supported by Natural Science Foundation of China(No. 10602027)
Baoyin Hexi: Associate Professor, Department of Aerospace Engineering, Tsinghua University.
Email: baoyin@tsinghua.edu.cn. Tel:62795211

Lagrange point $L_3$; see Fig.1 or Fig.2, case $C_2>C>C_3$). Therefore, when the comet enters Jupiter's influence region, it can only travel through $L_1$ and $L_2$, two necks of the zero velocity surfaces. Inside Jupiter's influence region it reflects on the zero velocity surfaces to become a temporary satellite of Jupiter. Might this also be the case with certain objects and Earth?

To date, researchers have found thousands of Near Earth Objects and continue to find more every week. While these NEOs do pose the threat of an impact with Earth, they also provide us with great opportunities. A 2*km*-size metallic NEO, for example, may contain rich metals and materials worth more than 25 trillion dollars (Hartmann et al. 1994). The concept of mining NEOs is not new (Gaffey et al. 1977), but there is still no proper practical way to do it. If approaching NEOs could be temporarily captured by Earth, exerting a small velocity change in the capture phase to bring them into orbit around Earth and finding a low-cost trajectory to sample a large amount of material would be one of the best ways to mine the NEOs. To deflect NEOs which are hazardous to Earth, different schemes have been presented, such as direct impact, mass driver, nuclear explosion, thrusting manoeuvres, and solar radiation (Ahrens et al. 1992; McInnes 2004; Chapman 2004; Ivashkin et al. 1995). All these schemes can be used for changing the orbital elements of the NEOs, but these schemes have only been examined with regard to NEOs of less than one meter in size; most Near Earth Objects are quite large and in high-energy orbit.

Many authors have studied the gravitational capture phenomenon by using different models of celestial mechanics (Makó et al. 2004). Brunini et al. examined the conditions of capture in the restricted three-body problem. Murison studied the connections between gravitational capture and chaotic motions. Makó and Szenkovits gave some necessary conditions of not being captured by using the Hill-regions in the spatial elliptic restricted three-body problem.

In this paper, we studied the necessary conditions of capture and identified how these conditions could lead to certain kinds of NEOs being captured by the Earth. Like Oterma being captured by Jupiter, it is possible for low-energy NEOs to be temporarily captured by the Earth; moreover, it is possible for them to become Earth-orbiting objects after the exertion of a small velocity increment. In our discussions of the orbital dynamics of the NEOs, we used the framework of an accurate solar system model when the NEOs were far from any major celestial body. The Sun-Earth-NEO three-body system was utilized when the NEOs were approaching the Earth. From the perspective of the restricted three-body system, it is possible for an NEO whose Jacobian constant is slightly lower than Sun-Earth $C_1$ and higher than $C_3$ to be temporarily captured by the Earth. When such an NEO approaches the Earth, it is possible to change its orbit energy to close up the zero velocity surfaces at point $L_1$ (see Fig.1 or Fig.2, case $C=C_1$; it will remain inside the smaller ball) and make it become a small satellite of the Earth. However, based on current technology, this may only possible with very small NEOs, as larger ones are too heavy for their orbit energy to be significantly changed. Fortunately, a practical advantage of such low-energy NEOs is that mining them requires less fuel and time than larger NEOs.

**2. System model**

**Orbit prediction:** Because of the perturbations, especially the resonance from the planets, the dynamical model only consists of Sun-Earth-NEO three body system that cannot accurately predict close approach of the near Earth objects. Fortunately there is some software that can be used for predicting NEO orbit very accurately in an appropriate time scale. Milani et al. have developed a very accurate solar system model, in which the gravitational forces of all bigger celestial bodies (including the Earth and the moon, which are treated as individual bodies instead of the barycentre of Earth-Moon system) of the solar system even the bigger NEOs themselves have been taken into account (Milani et al. 2001). That model is implemented in their free software package OrbFit and its source codes also can be freely downloaded from their website. This paper will use this software to calculate the NEO orbit and to predict their close approaches.

**The elliptical restricted three body model:** When we consider the orbit change of an NEO as it close approaches to the Earth, the system can be seen as the Sun-Earth-NEO elliptical 3-dimensonal three-body system. In such a system, the Sun and the Earth revolve around their common mass center in a Keplerian elliptical orbit under the influence of their mutual gravitational attraction. The NEO, of infinitesimal mass, moves in the 3-dimensional space under their gravitational influences.

To succinctly describe the elliptic restricted three-body problem, a non-uniformly rotating and pulsating coordinate system is used here. In this system, origin of the coordinates is in the common mass center of the two massive primaries, and the $x$ axis is directed towards $m_1$ (the Earth), the $xy$ plane is the orbit plane of the two massive primaries. Such a pulsating or oscillating coordinate system might be transformed to dimensionless coordinates by using the variable distance between the primaries as the length unit and the reciprocal of the variable angular velocity of the Earth as the time unit.

$$[L] = r = \frac{a(1-e^2)}{1+e\cos f}$$

$$[T] = \frac{1}{\dot{f}} = \frac{R^2}{\sqrt{G(m_1+m_2)a(1-e^2)}}$$

where $R$ is the mutual distance, $a$ and $e$ are the semimajor axis and eccentricity of the elliptic orbit, and $f$ is the true anomaly.

In the non-uniformly rotating and pulsating coordinate system, the two massive primaries are always in fixed location on the $x$ axis, and the dimensionless angular velocity of the two primaries is 1.

Consider first the equations of motion in inertial coordinate system, using dimensional quantities and variables the equations of motion can be given as (Szebehely et al. 1967)

$$\frac{d^2X}{dt^2} = -Gm_1 \frac{X-X_1}{R_1^3} - Gm_2 \frac{X-X_2}{R_2^3} \tag{1a}$$

$$\frac{d^2Y}{dt^2} = -Gm_1 \frac{Y-Y_1}{R_1^3} - Gm_2 \frac{Y-Y_2}{R_2^3} \tag{1b}$$

$$\frac{d^2Z}{dt^2} = -Gm_1 \frac{Z-Z_1}{R_1^3} - Gm_2 \frac{Z-Z_2}{R_2^3} \tag{1c}$$

where $t$ is the dimensional time, and $X$, $Y$ and $Z$ are the dimensional coordinates of the third body in inertial coordinate system,

$$R_i^2 = (X-X_i)^2 + (Y-Y_i)^2 \tag{1d}$$

$i = 1, 2$, and $X_i, Y_i$ are the dimensional coordinates of the two massive primaries.

The dimensional rotating coordinates $(\tilde{x}, \tilde{y}, \tilde{z})$ and the inertial coordinates $(X, Y, Z)$ satisfy

$$\begin{pmatrix} \tilde{x} \\ \tilde{y} \\ \tilde{z} \end{pmatrix} = \begin{pmatrix} \cos f & \sin f & 0 \\ -\sin f & \cos f & 0 \\ 0 & 0 & 1 \end{pmatrix} \begin{pmatrix} X \\ Y \\ Z \end{pmatrix} \quad (2)$$

The relationship between the dimensionless coordinates $(x, y, z)$ and the dimensional coordinates $(\tilde{x}, \tilde{y}, \tilde{z})$ can be given as

$$\begin{pmatrix} x \\ y \\ z \end{pmatrix} = \begin{pmatrix} \tilde{x} \\ \tilde{y} \\ \tilde{z} \end{pmatrix} / [L] = \begin{pmatrix} \tilde{x} \\ \tilde{y} \\ \tilde{z} \end{pmatrix} / \frac{a(1-e^2)}{1+e\cos f} \quad (3)$$

The true anomaly as the independent variable may be introduced by the equation

$$\frac{d}{dt} = \frac{d}{df}\frac{df}{dt} = \frac{d}{df} \cdot \dot{f} \quad (4)$$

Substitute equation (2), (3), (4) into the equation (1a), (1b) and (1c) and let

$$\mu = \frac{m_2}{m_1 + m_2} \quad (5a)$$

$$1 - \mu = \frac{m_1}{m_1 + m_2} \quad (5b)$$

where $\mu$ is the dimensionless mass of the Earth and $1-\mu$ then the dimensionless mass of the Sun. In addition, the dimensionless $x$ coordinate of the Sun is $\mu$ and that of the Earth is $1-\mu$.

In the rotating frame, the two primaries are fixed and dimensionless equations of NEO motion in a non-uniformly rotating and pulsating coordinate system can be obtained as

$$\ddot{x} - 2\dot{y} = \frac{\partial \Omega}{\partial x} \quad (6a)$$

$$\ddot{y} + 2\dot{x} = \frac{\partial \Omega}{\partial y} \tag{6b}$$

$$\ddot{z} + z = \frac{\partial \Omega}{\partial z} \tag{6c}$$

where the dots imply the derivative with respect to true anomaly $f$ of the Earth, and

$$\Omega = (1 + e\cos f)^{-1}\left(\frac{x^2 + y^2 + z^2}{2} + \frac{1-\mu}{r_1} + \frac{\mu}{r_2}\right) \tag{7a}$$

$$r_1^2 = (x + \mu)^2 + y^2 + z^2 \tag{7b}$$

$$r_2^2 = (x - 1 + \mu)^2 + y^2 + z^2 \tag{7c}$$

Similar to the circular restricted three-body problem, differential equation (6) possesses some important properties that are its local five equilibrium points, three collinear points and two equilateral points and the zero velocity surfaces.

**The local Jacobian constant and the local zero velocity surfaces:** The invariant quantity, well-known Jacobian integral, which exists in the circular case, does not exist globally in this case. Here, however, let's define the local Jacobian constant and the local zero velocity surfaces. Multiplying Eq. (6a) by $\dot{x}$, (6b) by $\dot{y}$ and (6c) by $\dot{z}$, adding them, and integrating, it yields

$$\dot{x}^2 + \dot{y}^2 + \dot{z}^2 = 2\Omega - 2e\int \frac{\sin f}{1 + e\cos f}\Omega df - C - z^2 \tag{8}$$

If one considers a short time arc of a near circular orbit, in Eq. (8) the integral term can be negligible, therefore the local Jacobian constant can be defined as

$$C = 2\Omega - v^2 - z^2 \tag{9}$$

The Eq. (8) implies that the 3-dimensional elliptical three-body problem does not have global zero velocity surfaces. Therefore the local zero velocity surfaces can be defined as

$$C = 2\Omega - z^2 \tag{10}$$

The shape of the zero velocity surfaces depending on the value of constant $C$ are shown as in Fig.1, similar to the figures shown in ref. [8].

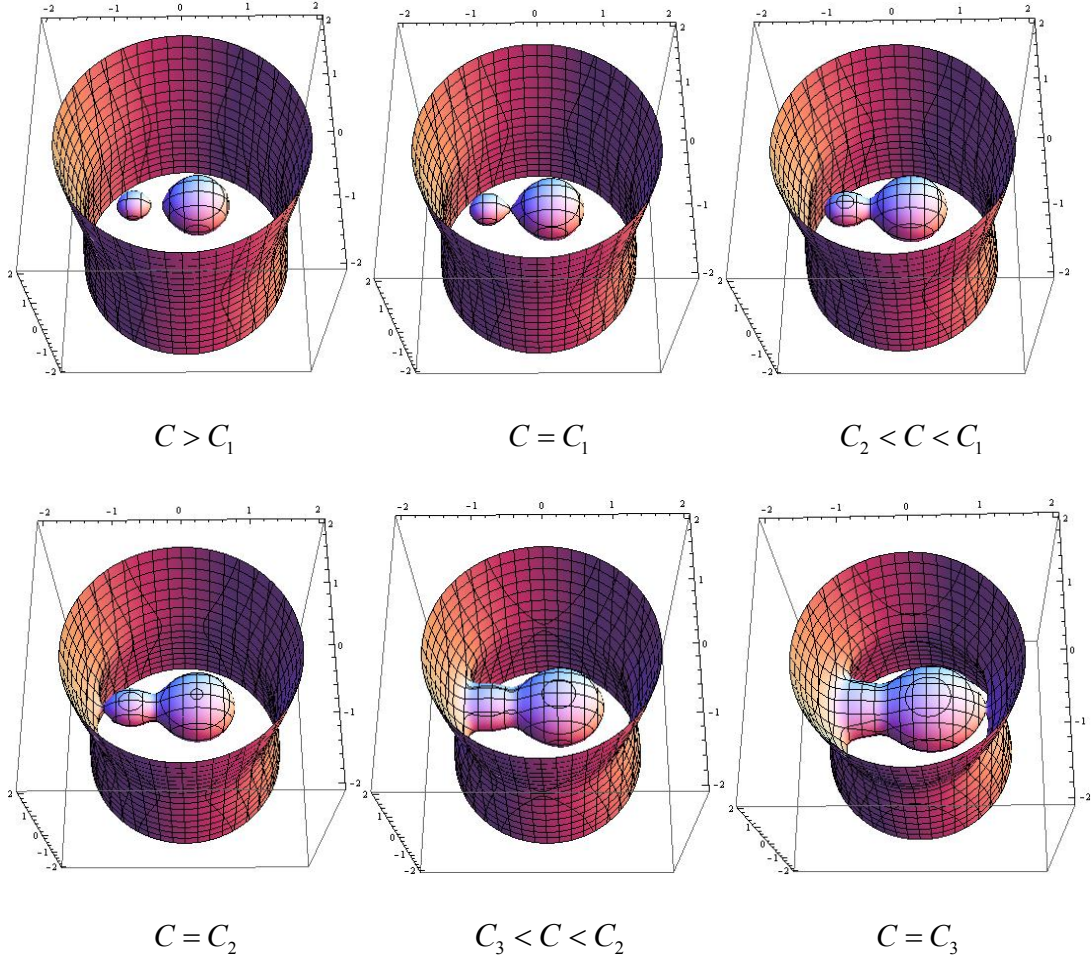

$C > C_1$     $C = C_1$     $C_2 < C < C_1$

$C = C_2$     $C_3 < C < C_2$     $C = C_3$

Fig.1 the zero velocity surfaces for different value of $C$

The zero velocity surfaces form the boundaries between forbidden and allowed regions of motion. The planar ($z=0$) allowable regions of motion (Hill-zone) for different values of Jacobian constant are shown in Fig.2 (shaded). Here we define the Earth influence region as the allowed regions of motion (Hill-zone) surrounding the Earth when $C=C_1$ and illustrated it in Fig.2a (to show clearly the Fig.2 here took $\mu=0.2$ rather than the $\mu$ of the Sun-Earth system)

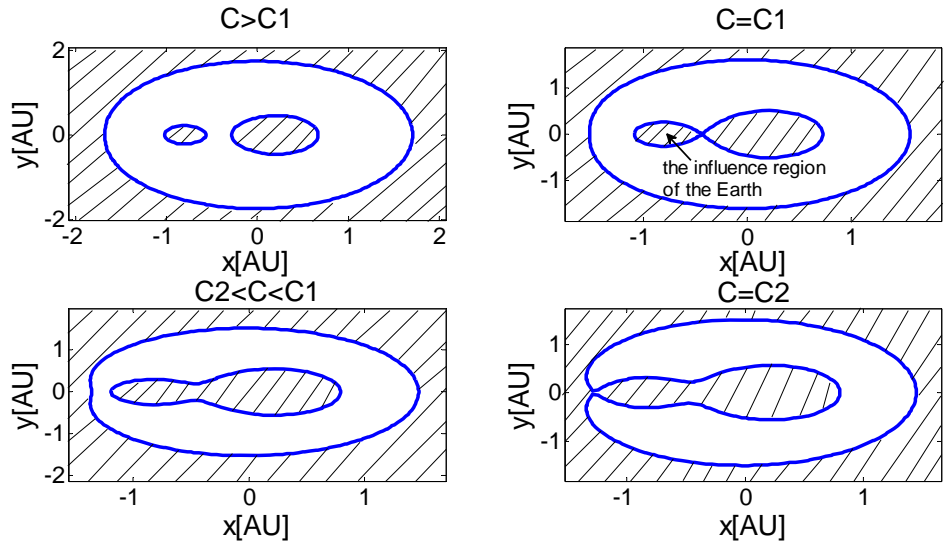

Fig.2a the allowable regions of motion for different values of $C$ ($\mu$=0.2)

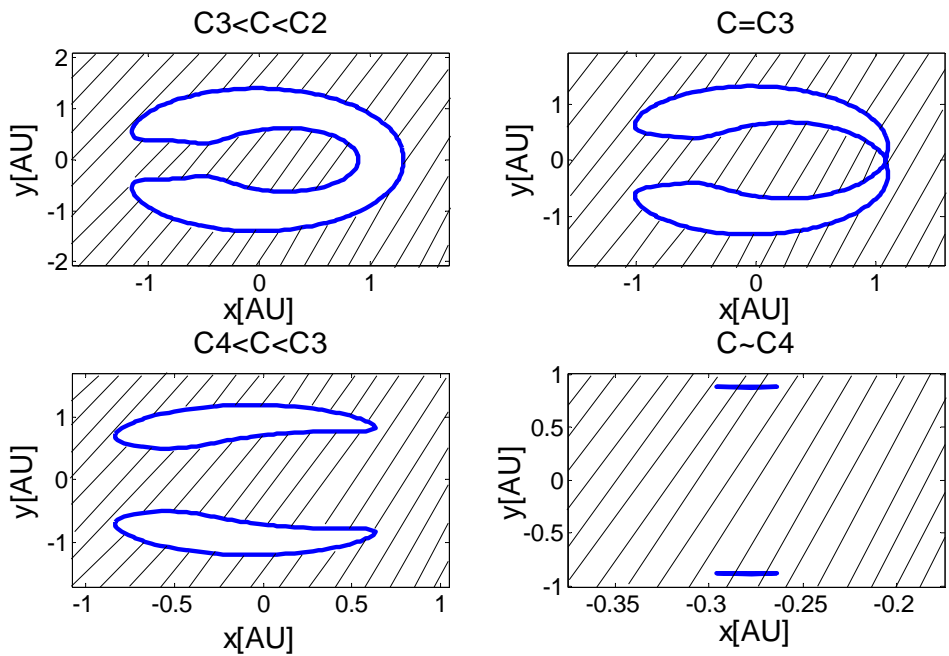

Fig.2b the allowable regions of motion for different values of $C$ ($\mu$=0.2)

**Necessary conditions of the capture:** This zero velocity surfaces can provide a necessary condition of temporal capture (Because the NEOs also affected by the gravitational pull of other celestial bodies except the Sun and the Earth, So only the condition of temporal capture can be given). If an NEO is inside the zero velocity surfaces surrounding the Earth, and satisfying the condition $C>C_1$, then it would not penetrate the zero velocity surfaces to become a small satellite of the Earth.

We use this property to search the possible temporal captured NEOs and to estimate the velocity requirement for changing an NEO orbit into the orbit around the Earth when a NEO enters the Earth influence region.

We added some codes including sequence control function to the software ORBFIT and examined all the close approaching NEOs (considered the Earth approaching distance is closer than $0.008AU$ and time is up to year 2060) from over six thousand NEOs[*]. The local Jacobian constants of the close approaching NEOs are listed in Table 1. From table 1 we can conclude that there are no close approaching NEOs to seem naturally captured by the Earth temporarily. But the Jacobian constant of some NEOs close to $C_1$, such as 2008EA9 and 2009BD, others are quite different from it. But we cannot deny that we may find some such kind of NEOs in the future.

### 3. Orbital change of the close approaching NEO

**Orbital change of an NEO:** Here we consider giving a velocity increment to change NEO orbit energy to have it captured by the Earth. In the close approaching phase, the orbital dynamics is considered as Sun-Earth-NEO elliptical three-body system, so using the local Jacobian integral, it can be written as

$$v^2 = 2\Omega - C + z^2 \qquad (12)$$

If a velocity increment is applied in a short time, the velocity difference can be written as

$$v_1^2 - v_0^2 = 2\Delta\Omega - \Delta C + \Delta z \qquad (13)$$

---

[*] The orbital data of these asteroids can refer to http://neo.jpl.nasa.gov/cgi-bin/neo_elem

The maneuver is assumed to be implemented in a short time, than ΔΩ=0, Δz=0 can be approved, therefore

$$\Delta v = \sqrt{v_0^2 - \Delta C} - v_0 \quad (14)$$

For engineering application, this velocity increment is not only a criterion for local capture, but also a useful value for NEOs sampling. For a large amount of NEO sample, for example, choosing a small velocity increment NEO can cut the fuel cost of spacecraft greatly. Table 1 shows that most of the close-approaching NEOs are in the high energy orbit, see the column of *C* and Δ*V*. It is very difficult to change these high energy orbit NEOs to have them captured by the Earth, but there are some NEOs which orbital energy is easy to be changed, like last two of the table.

Table 1. The Jacobian constant of the close approaching NEOs (the parameters are referenced from NEO Information Services of Pisa University and JPL Near Earth Object Program)

| Name | Abs Mag | Dia (m) | Approach(AU) (year/month) | Local C | Δ$v$ (km/s) |
|---|---|---|---|---|---|
| 137108 | 17.9 | 800 | 0.002607 (2027/08) | 2.25105 | -24.82 |
| 2008KO | 17.4 | 30-80 | 0.00753075(2052/06) | 2.41202 | -22.66 |
| 2002NY40 | 19.0 | 280 | 0.00730030(2038/02) | 2.47592 | -20.37 |
| 2007EH26 | 24.2 | 40-90 | 0.00498466(2049/09) | 2.57581 | -18.87 |
| 2007PF2 | 24.4 | 30-80 | 0.00512391(2024/08) | 2.62260 | -18.14 |
| 221455 | 19.4 | 370-840 | 0.00783329(2052/3) | 2.62078 | -17.55 |
| 2009SU104 | 25.6 | 20-50 | 0.00147771(2057/02) | 2.66291 | -14.74 |
| 2000LF3 | 21.6 | 140-320 | 0.0079286(2046/06) | 2.78643 | -14.42 |
| 2004VZ14 | 25.3 | 20-50 | 0.00596776 (2042/11) | 2.73614 | -14.06 |
| 2008YF | 20.9 | 190-440 | 0.00362470(2035/12) | 2.70749 | -13.96 |
| 1998HH49 | 21.3 | 280 | 0.0078544(2023/10) | 2.76691 | -13.46 |
| 2007VX83 | 27.8 | 0-10 | 0.00555639(2030/11) | 2.76124 | -13.09 |
| 2000YA | 23.7 | 700 | 0.00735641(2011/12) | 2.74495 | -13.07 |
| 2005EQ95 | 23.4 | 60-140 | 0.00708734(2045/10) | 2.78439 | -12.92 |
| 1997XF11 | 16.7 | 1300-2800 | 0.00621084(2028/10) | 2.77705 | -12.87 |
| 2009RR | 25.6 | 20-50 | 0.00713957(2014/09) | 2.81182 | -12.74 |
| 2001TB | 24.8 | 30-80 | 0.00482637(2010/10) | 2.80398 | -12.19 |
| 2005YU55 | 22.0 | 120-280 | 0.00362956(2011/11) | 2.78383 | -12.10 |
| 2007VL3 | 26.0 | 10-40 | 0.00740713(2014/10) | 2.82102 | -11.34 |
| 2009WV25 | 24.0 | 40-100 | 0.00644959(2053/04) | 2.86257 | -10.71 |
| 2008GY21 | 27.6 | 0-20 | 0.00444996 (2018/04) | 2.85320 | -10.67 |
| 2007TL16 | 26.2 | 10-30 | 0.006224(2037/10) | 2.86095 | -10.29 |
| 2006BX147 | 25.8 | 150-350 | 0.00205264(2013/01) | 2.89963 | -9.79 |
| 2004VZ | 24.5 | 30-80 | 0.00414594 (2056/11) | 2.83855 | -9.96 |
| 2008VM | 30.2 | 0-0 | 0.00474005(2051/11) | 2.84631 | -9.86 |
| 2006DM63 | 26.7 | 10-30 | 0.00524844(2053/02) | 2.84449 | -9.79 |
| 2008EX5 | 23.8 | 50-110 | 0.00428753(2042/10) | 2.87449 | -9.46 |
| 2009EU | 26.6 | 10-30 | 0.00443390(2043/03) | 2.86389 | -9.27 |
| 153814 | 18.2 | 700 | 0.0016667(2028/06) | 2.93194 | -8.45 |
| 2007UT3 | 25.7 | 10-40 | 0.00753056(2028/08) | 2.94368 | -8.06 |
| 2008LH2 | 24.4 | 30-80 | 0.00640139(2039/06) | 2.95929 | -7.87 |
| 2009TM8 | 28.6 | 0-10 | 0.00283001(2011/10) | 2.91328 | -7.23 |
| 2007JB21 | 25.4 | 20-50 | 0.00649834(2054/05) | 2.96018 | -6.86 |
| 2004UT1 | 26.4 | 10-30 | 0.00770637 (2022/10) | 2.92514 | -6.38 |
| 162162 | 19.7 | 160-370 | 0.0065285 (2048/02) | 2.9226 | -6.15 |
| 2008TC3* | - | 0-0 | 0.00003910 (2008/10) | 2.98029 | -2.04 |
| 2005VL1 | 26.7 | 10-20 | 0.00558563(2049/02) | 2.91503 | -5.4 |
| 2008GM2 | 28.4 | 0-10 | 0.00538838(2035/04) | 2.96996 | -4.3 |
| 2008LG2 | 25.2 | 20-60 | 0.00673584(2056/06) | 3.02363 | -4.17 |
| 2005TA | 27.2 | 10-20 | 0.00297955(2034/10) | 2.97108 | -4.15 |
| 2005TA | 27.2 | 10-20 | 0.00265873 (2041/10) | 2.97172 | -4.1 |
| 2009QR | 27.3 | 0-20 | 0.00235132(2023/08) | 3.00272 | -4.03 |
| 2009WR52 | 28.3 | 0-10 | 0.00303283(2028/05) | 3.00523 | -4.03 |
| 2006HE2 | 26.5 | 10-30 | 0.00518549(2029/09) | 2.98176 | -3.76 |
| 2003LN6 | 24.5 | 30-70 | 0.00672417(2053/04) | 2.99763 | -3.66 |
| 2006WB | 22.8 | 80-180 | 0.00595314(2024/11) | 2.93867 | -3.64 |
| 2007XB23 | 27.1 | 10-20 | 0.00292837(2024/12) | 2.93334 | -3.53 |
| 2008HQ3 | 18.1 | 0-20 | 0.00717914(2038/04) | 2.98815 | -3.3 |
| 2001AV43 | 24.4 | 30-70 | 0.00742242(2013/11) | 2.95181 | -3.24 |
| 2007UD6 | 28.3 | - | 0.00026283(2048/10) | 2.95841 | -2.59 |
| 99942 | 19.2 | 270 | 0.00025418(2029/4) | 2.97612 | -2.39 |
| 2008EA9 | 27.7 | 0-10 | 0.00694714(2049/02) | 2.95187 | -1.0 |

---

* According to our calculations, on Oct 2008 the minimum distance between 2008TC3 and the Earth is 0.00003910AU, which is less than the radius of the Earth. In fact, this object has been hit the Earth and exploded over the Sudan on 7 Oct 2008.

| | | | | | |
|---|---|---|---|---|---|
| 2009BD | 28.3 | 0-10 | 0.00231372(2011/06) | 3.03807 | -0.67 |
| 2009BD | 28.3 | 0-10 | 0.00458782(2009/01) | 2.95112 | -0.41 |

The NEOs' diameter data listed in the Table 1 can be computed using the following equation (Steven et al. 2002)

$$D = \frac{1329}{\sqrt{P}} \times 10^{-0.2H}$$

where $H$ is the absolute magnitude of the NEO, $P$ is the geometric albedo of the NEO, $D$ is the diameter of the NEO in kilometers.

It should be addressed that the orbital data we adopted are not so exact, because due to the limitation of the observations some orbits of NEOs are poorly determined.

It is obvious from Eq. (12) that the Jacobian constant is actually not a constant in the elliptical system, but it is varying with true anomaly of the Earth orbit. However, when the true anomaly is fixed at a close approaching moment, one can get a local Jacobian constant. Fig.3 shows the time varying Jacobian constant of the close approaching NEOs vs Sun-Earth $C_1$ at the corresponding true anomaly.

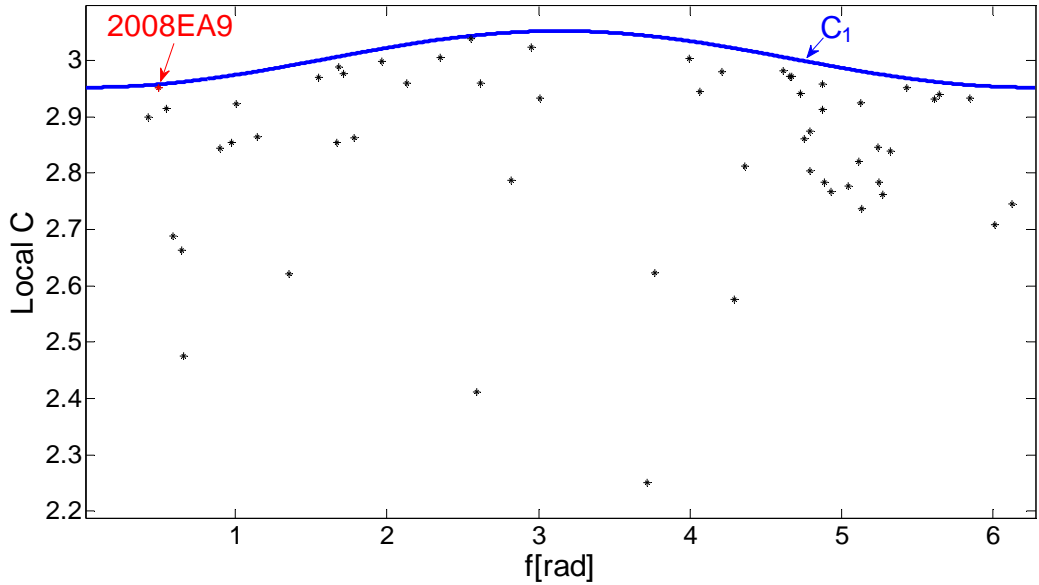

Fig.3. the local *C* of all the NEOs (Earth approaching distance ≤ 0.008AU)

## 4. Capture example

**2008EA9:** 2008EA9 is a 10-metre near Earth object, the orbital data of it is shown in Table 2. And Fig.4 shows the orbits of the Earth and the NEO 2008EA9 in the Sun centered coordinates. From Fig.3 and Table 1, it can be seen that the velocity increment of the 2008EA9 is relatively small (-1.00*km/s*) and it will very close approach (0.00694761*AU*) to the Earth in 2049/02. Moreover the size of the NEO 2008EA9 is very small so that the capturing of it is relatively easy.

Table 2. The orbital data of the NEO 2008EA9 (epoch: MJD 55200)

| *a*(*AU*) | *e* | *i*(*deg*) | Ω(*deg*) | ω(*deg*) | *M*(*deg*) |
|---|---|---|---|---|---|
| 1.05916 | 0.07978 | 0.424 | 129.426 | 335.944 | 298.104 |

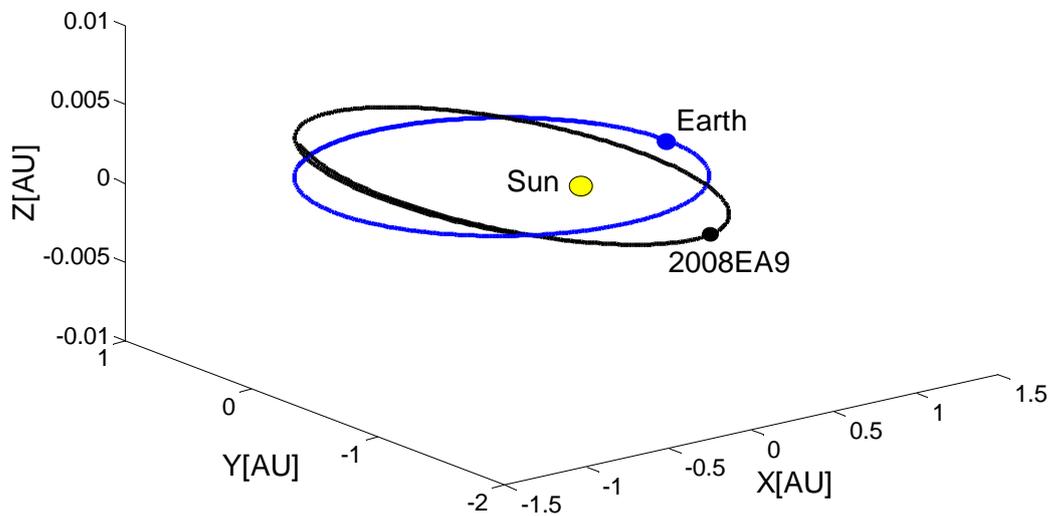

Fig.4. the orbits of the Earth and the NEO 2008EA9

We simulated the trajectory before and after the orbit maneuver using the accurate dynamic model, the result is shown in Fig.5. Fig.5 shows that the orbit of 2008EA9 after maneuver is very close to the Earth orbit in *xy* plane, only existing $10^{-3}AU$-order's pulsation in *z* axis. Fig.6 shows the trajectory of the NEO 2008EA9 after maneuver in a geocentric inertial coordinate system. After the orbit maneuver, the NEO becomes an Earth temporal orbiting satellite with orbit altitude about $0.005AU$, a distance twice of the Earth-Moon distance.

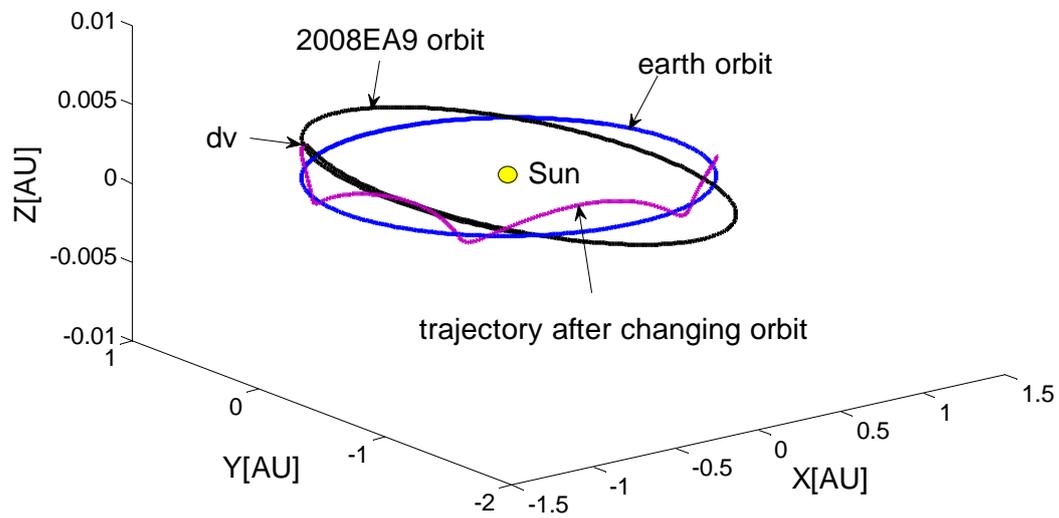

Fig.5a trajectory of 2008EA9 before and after orbit maneuver in a Sun-centred coordinates

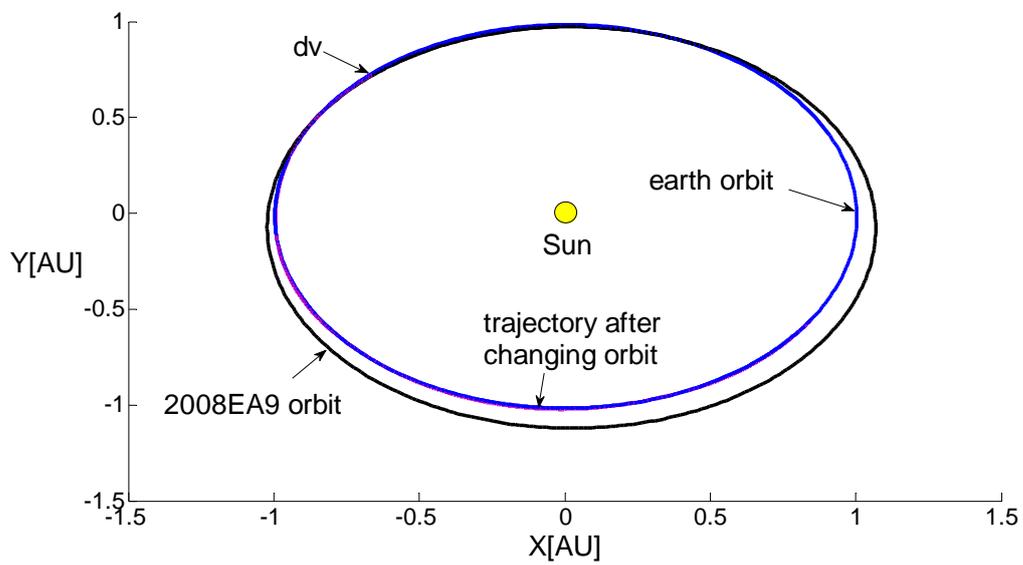

Fig.5b trajectory of 2008EA9 before and after orbit maneuver in a Sun-centred coordinates (*xy*)

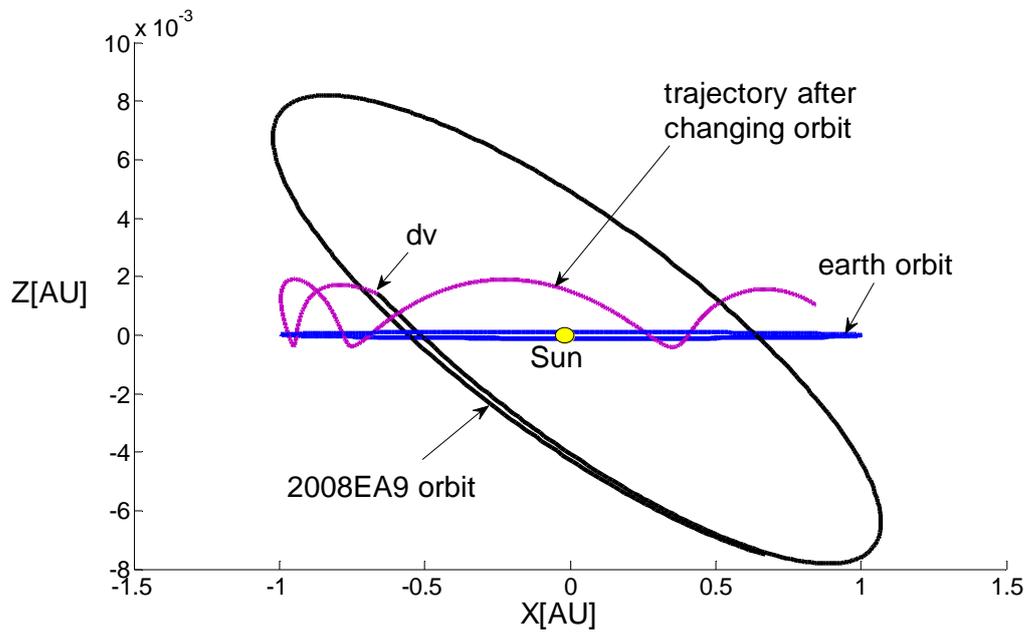

Fig.5c trajectory of 2008EA9 before and after orbit maneuver in a Sun-centred coordinates (*xz*)

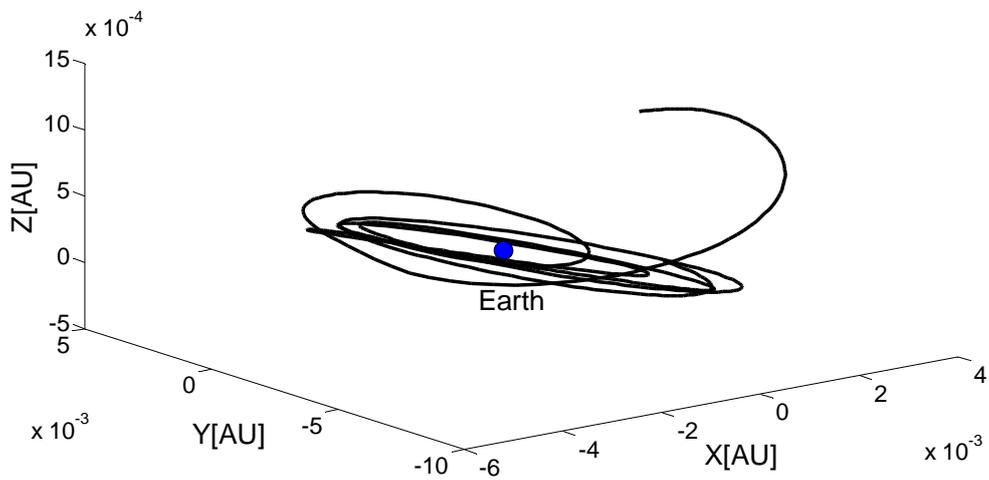

Fig.6a trajectory of 2008EA9 after orbit change in geocentric inertial coordinates

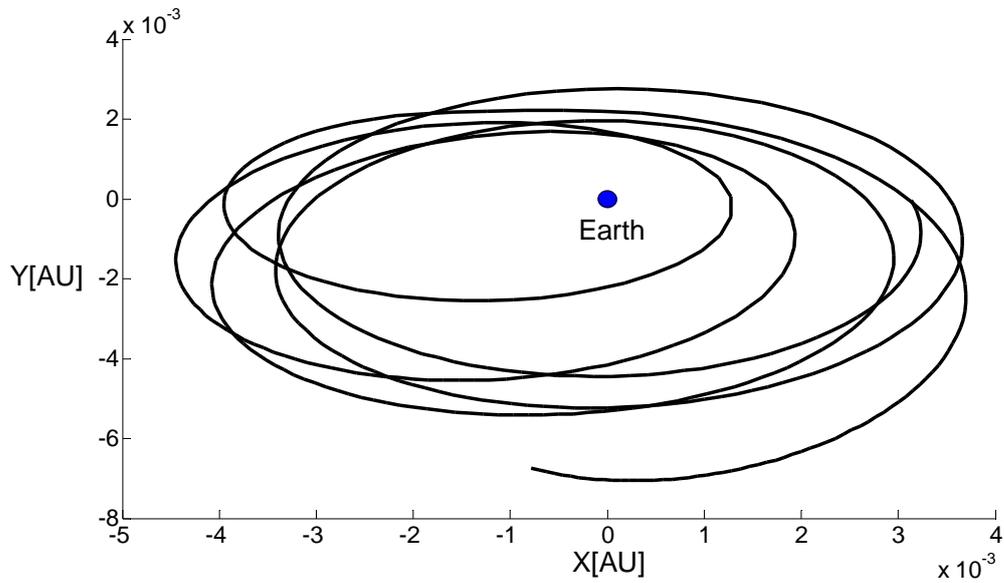

Fig.6b trajectory of 2008EA9 after orbit change in geocentric inertial coordinate (*xy*)

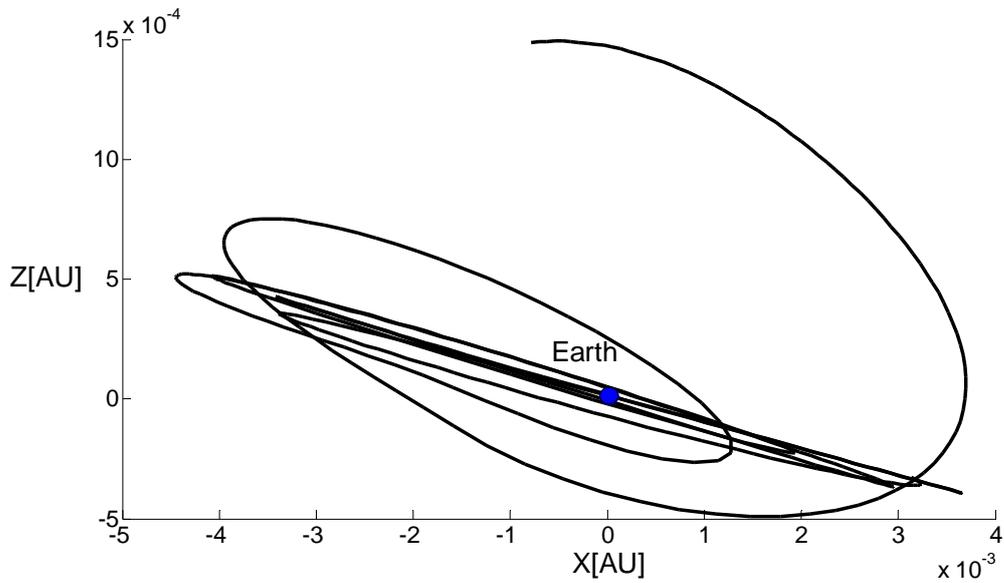

Fig.6c trajectory of 2008EA9 after orbit change in geocentric inertial coordinate (*xz*)

## 5. Capture methods

In order to capture a near Earth object, there are several alternatives. These alternatives were broadly classified as "impulsive" if they acted nearly instantaneously, or "slow push" if they acted over an extended period of time (NASA 2006). The impulsive techniques generally included conventional explosive, kinetic impactor and nuclear explosive. And the slow push techniques included Enhanced Yarkovsky effect, focused solar, gravity tractor, mass driver, pulsed laser and space tug. Considering the required impulsive velocity increment is not so small and the diameter of NEOs is relatively large, there are two impulsive capture methods available, kinetic impactor and nuclear explosion, but they are never tested or applied. Among them, the nuclear explosion method may not be proper one for the mentioned small NEO, because the nuclear explosion can release a very large amount of energy, the result may be a fragmentation of the target NEO. So the kinetic impactor is often considered as a better maneuver means especially for the NEOs smaller than 50 meters in diameter.

**Kinetic impact:** A space probe or a specially designed projectile which will hit an NEO at a very high velocity, and therefore deliver an impulse that will change the orbit of the NEO. The relative velocity between the impactor and the NEO depends on the orbit of the impactor, the limited payload capacity of the available launch systems, and the application of new technology. For example, the relative velocity can reach 60km/s by utilizing solar sailing (McInnes 2004).

For the kinetic impactor, the transfer of momentum can be calculated by

$$\Delta V = \beta \frac{m_i v_i}{M_a} \tag{15}$$

where $\beta$ is the impact efficiency constant, the value of the constant greatly depends on the structure and the material properties of the NEO. If $\beta=1$, the collision is perfectly plastic, no ejecta are produced, and momentum is imparted directly. If an NEO is sufficiently soft and the impactor penetrates, then $\beta<1$ and the impact is less effective. If that $\beta>1$, ejecta are released by the impact, and the impact is more effective. Some estimate that $\beta$ could have a magnitude of 10 or more.

The effective momentum changes for kinetic impact are calculated to range from about $5\times10^7$ to $2\times10^{10}$ kg-m/s (NASA 2006). Taking the NEO 2008EA9 as an example, capturing it needs a velocity increment of -1.00$km/s$, if its density is $2\times10^3 \, kg/m^3$ and diameter is 10$m$, the momentum changes can be given as

$$\Delta p = M\Delta v = \rho V \Delta v = 8\times10^9 \, kg\cdot m/s \qquad (16)$$

So the kinetic impact is a possible way for the capture of 2008EA9, and assuming the relative velocity $v_i$=60$km/s$ and the impact constant $\beta$=5, we can calculate the mass of impactor needed by Eq.15

$$m_i = \frac{\Delta p}{v_i \beta} = 26.4\times10^3 \, kg \qquad (17)$$

This is still a likely practical value.

**Sampling from a close approaching NEO:** Large amount sampling desires low fuel cost trajectory, however, the near Earth objects are not all in the desirable energy orbit. Mining some NEOs needs more velocity increment while others need lower. But if it is possible, mining in close approaching moment can greatly shorten mission time that may significantly reduce risk. Here we assume that mining an NEO and depart from the NEO at its closest point of the Earth, and in this point the spacecraft changes its velocity so that into a big elliptical Earth centred orbit whose perigee intersect the parking orbit (for example 200$km$), and after reaches the perigee applying another velocity change to get parking orbit. The $L_1$ closing velocity increment can be obtained by Eq. (14). And according to the orbital energy equation, in the apogee and perigee the velocity increments can be conservatively estimated by

$$\Delta v_a = \sqrt{\frac{\mu_e}{r_a}}\left[1-\sqrt{\frac{2r_p}{(r_a+r_p)}}\right] \qquad (18a)$$

$$\Delta v_p = \sqrt{\frac{\mu_e}{r_p}}\left[\sqrt{\frac{2r_a}{(r_a+r_p)}}-1\right] \qquad (18b)$$

Where $\mu_e$ is the gravitational constant of the Earth; $r_a$ is the apogee radius and $r_b$ is the perigee radius, respectively, of Earth centred orbit. The transfer mentioned above similar to a Hohmann transfer between a big and a small circular orbit. One can obtain $\Delta v_a$=0.73$km/s$ and $\Delta v_p$=2.19$km/s$ for NEO 2008EA9. Counting $L_1$ closing velocity increment, it totally needs velocity increment of 3.92$km/s$ for returning from the NEO, and it would be much lower if the aerobraking is considered.

**6. Conclusions**

In this paper, we examined Near Earth Objects' orbital dynamics using the framework of an accurate solar system model and the Sun-Earth-NEO three-body system to search for low-energy NEOs that may be temporarily captured by the Earth or might be able to orbit the Earth after the exertion of a small velocity increment. The results showed that none of the more six thousand NEOs were naturally captured, but we did find some NEOs that would be captured by the Earth after exerting a small velocity change (less than 1km/s) while close to the Earth. These NEOs are prime candidates for short sampling missions conducted by spacecraft.